\documentstyle[12pt,psfig]{article}
\newcommand{\be}{\begin{equation}}
\newcommand{\ee}{\end{equation}}
\newcommand{\bea}{\begin{eqnarray}}
\newcommand{\eea}{\end{eqnarray}}
\newcommand{\grgl}{\:\hbox to -0.2pt{\lower2.5pt\hbox{\small$\sim$}\hss}
                 {\raise3pt\hbox{$>$}}\:}
\newcommand{\klgl}{\:\hbox to -0.2pt{\lower2.5pt\hbox{\small$\sim$}\hss}
                 {\raise3pt\hbox{$<$}}\:}
\begin{document}

\begin{center}
{\Large \bf Fluctuations of the transverse energy in {\boldmath
	 $Pb\!+\!Pb$} collisions and {\boldmath $J/\psi$}
	 suppression}\\[5mm] 
{J\"org H\"ufner$^{a,b}$, 
Boris Z. Kopeliovich$^{b,c}$ and 
Alberto Polleri$^a$\footnote{{\it Present address}: Physik Department, 
TU M\"unchen, James-Franck-Stra\ss e, D-85747 Garching, Germany. 
e-mail: polleri@physik.tu-muenchen.de}}\\[4mm]
	 {\small{\it $^a$ Institut f\"ur Theoretische Physik der
	 Universit\"at, Philosophenweg 19, D-69120 Heidelberg,
	 Germany.} \\[1mm] {\it $^b$ Max Planck Institut f\"ur
	 Kernphysik, Postfach 103980, D-69029 Heidelberg, Germany.}
	 \\[1mm] {\it $^c$ Joint Institute for Nuclear Research,
	 Dubna, 141980 Moscow Region, Russia.} \\[5mm]}
\today\\[5mm]
\end{center}

\begin{abstract}
\setlength{\baselineskip}{16pt}
The observed $J/\psi$ suppression in $Pb+Pb$ collisions shows a drop
at those large values of the transverse energy $E_T$ which arise from
fluctuations. The validity of existing models for $J/\psi$ suppression
can be extended into this domain of $E_T$ by introducing an {\it
ad hoc} factor proportional to $E_T$. We propose a formalism in which
the influence of $E_T$ fluctuations on the $J/\psi$ suppression can be
calculated and discuss the conditions under which the {\it  ad hoc}
factor is obtained.
\end{abstract}

\vspace{0.3in}
 
\setlength{\baselineskip}{18pt}
 
The observed dependence on transverse energy of
$J/\psi$ production in $Pb\!+\!Pb$ collisions at 158$\,A$ GeV is the
first unambiguous signal for an anomalous mechanism for charmonium
suppression, {\it i.e.} one which goes beyond what is already observed
in proton-nucleus collisions and in reactions with light ions.
 In order to identify the detailed nature
of the anomalous mechanism, it is important to investigate each of its features
using a {\it minimum} of adjustable 
parameters. The
past experience has shown that models succeed to reproduce the
data only after several parameters are adjusted. 
Three years ago, the
NA50 collaboration has re-measured the $J/\psi$ suppression in
$Pb\!+\!Pb$ collisions in the region of large transverse energy
$E_T$ and has corrected its earlier result
\cite{NA50}. Rather than being flat
as a function of $E_T$ the new data show a drop
at $E_T \grgl 100$ GeV.  For an overview of the field, the
data and their interpretation we refer to the reviews \cite{GH,V} and
the proceedings of the  Quark Matter '99 conference \cite{QM}.
 
Within the picture of the quark-gluon plasma the newly observed drop
is interpreted as the expected onset of $J/\psi$ melting
\cite{S}.
 Two other groups, Capella {\it et al.} \cite{CFK} and Blaizot {\it
et al.} \cite{BDO}, modify their previously successfully proposed
expressions for the observed $J/\psi$ suppression by introducing the
factor
\be
\label{(1)}
\epsilon(E_T) = \frac{E_T}{\bar
E_T(b)}
\label{eps}
\ee
at the appropriate place, without the need of a new parameter. They
argue that the newly discovered drop for $J/\psi$ suppression
arises in the regime where the increase in $E_T$ is not caused by
a decrease in impact parameter $b$, but rather by fluctuations in $E_T$
around the mean value $\bar E_T(b)$, which is determined by the
collision geometry. Larger values of $E_T$ lead to a correspondingly
larger suppression. For the comover description by Capella {\it et
al.} the modification by the factor $\epsilon$ has  a small  effect
and fails to describe the data\footnote{In a recent paper
Capella {\it et al.} \cite{CKS} argue, that the drop shown in the
data is misleading, since the data are evaluated with respect to
minimum bias events, whose $E_T$ distribution differs from the $E_T$
distribution of events in which also a $\psi$ is observed.}, 
while the same modification to
the cut-off model by Blaizot {\it et al.} quantitatively describes the
newly observed break. These partial and full successes, respectively, 
call for a derivation
of the {\it ad hoc} formula. This is what is attempted in this paper.
 
The expression for the charmonium $(\psi)$ production cross section in $A+B$
collisions is usually  written as
\bea
\frac{d\sigma^{\psi}_{AB}}{dE_T}(E_T) & = & \sigma^{\psi}_{pp} 
\int d^2b\ P_T(E_T,b) \label{cross} \\
& & \times\, \int d^2s \ T_A\,T_B\ S_{nucl}(\vec b,\vec s)
\ S_{FSI}(\vec b,\vec s)\,, \nonumber
\eea
where
\be
S_{nucl} =
\frac{1 - \exp\left(-\sigma^{abs}_{\psi N}\,T_A\right)}
{\sigma^{abs}_{\psi N}\,T_A}
\ \frac{1 - \exp\left(-\sigma^{abs}_{\psi N}\,T_B\right)}
{\sigma^{abs}_{\psi N}\,T_B}
\ee
represents the part of the suppression (also present in $pA$
collisions) which is related to the propagation of the charmonium through both
nuclei, the $\psi$ being destroyed in $\psi N$ collisions with a rate
determined by the absorption cross section $\sigma^{abs}_{\psi N}$. 
We denote by
$T_A = T_A(\vec s)$ and $T_B = T_B(\vec b-\vec s)$ the
nuclear thickness functions. $P_T$ represents the probability of observation
of $E_T$ in events with impact parameter $b$, and is normalized to 1 when
integrated over $E_T$.
The anomalous part  $S_{FSI}$  accounts for final
state interactions with the produced quarks and gluons (QGP) or the
hadrons (comovers).
  
In the comover approach \cite{CFK} one writes
\be
S_{FSI}^{co}(\vec b,\vec s)=\exp\{-\sigma_{co} N_y^{co}(\vec b,\vec s)
\,\ln(N_y^{co}(\vec b,\vec s)/N_f)\}.
\label{com}
\ee
The suppression function depends on the density of comovers $N_y^{co}$
 at  the rapidity $y$ of the $\psi$, 
while $\sigma_{co}$ describes the $\psi$ absorption by a
comoving meson and is usually taken as the adjustable parameter of the
theory (of
the order of 1 mb). The corresponding density of comovers in $pp$
collisions is denoted by $N_f$.
 
In the cut-off model \cite{BDO} one assumes
\be
S_{FSI}^{cut}(\vec b,\vec s)=\Theta(n_p^c-n_p(\vec b,\vec s))\,,
\label{cut}
\ee
where the (mean) density  of
 participant nucleons in impact parameter
space is
\be
n_p(\vec b,\vec s) = T_A
\left[1-\exp(-\sigma_{pp}T_B)\right]+(T_A 
\leftrightarrow T_B)\,.
\label{part}
\ee
The ``critical participant density'' $n_p^c$, a parameter
adjusted to $n_p^c=3.7$~fm$^{-2}$,
represents the
density above  which $\psi$ absorption is $100\,\%$ effective.
The QGP phase transition may (but must not) be the mechanism for the 
critical transition.
 In ref. \cite{BDO} the
$\Theta$-function in eq.~(\ref{cut}) is  smeared at the
expense of a further parameter $\lambda$, which is obtained from a fit
to the data:
\be
S_{FSI}^{cut}(\vec b,\vec s)=\frac{1+\tanh[\,\lambda(n_p^c
-n_p(\vec b,\vec s))\,]}{2}.
\label{(7)}
\ee

In order to describe the endpoint behavior of charmonium suppression
at large $E_T$, the factor $\epsilon$  of eq.~(\ref{eps}) is
introduced in front of the factor $N_y^{co}$ in eq.~(\ref{com}),
with the argument that the number of comovers fluctuates proportionally to the
observed $E_T$, a very plausible assumption. In eq.~(\ref{cut})
for the cut-off model, the factor $\epsilon$ multiplies $n_p$, the
mean number of participants. This modification is not obvious. Why
should  the number of participants in the tube, where a $\psi$ is
produced, fluctuate proportionally to the global value of
$E_T$ ? For instance, in  a central collision practically all nucleons
participate.   Their number equals 
 $A\!+\!B$ and cannot  fluctuate. However, the
number of produced particles or the 
produced energy density does
fluctuate. Therefore, if the introduction of the factor $\epsilon$ in
front of $n_p$ has to make sense, we must interpret $n_p$ and
$n_p^c$ as being
{\it proportional} to the energy or particle 
density in the tube where $J/\psi$ is
produced. This is indeed the point of view of ref. \cite{BDO}. A
proportionality factor is irrelevant in the $\Theta$-function, but has
bearing on the parameter $\lambda$.
 
The actual situation is somewhat more complicated: the hadrons which are
observed as transverse energy are measured in a pseudo-rapidity
interval $1\leq\eta\leq 2.3$, while the $\psi$ 
is measured in the rapidity interval $3\leq y\leq 4$. (Recall that at the SPS
mid-rapidity corresponds to $y \simeq 3$.) It is not
obvious to which degree fluctuations  in $E_T$ in one rapidity interval should
have a bearing on the comovers  or  the energy
density which suppress the charmonium in another rapidity
interval. There must be a mechanism for ``cross talk''.

The paper addresses the following two issues:\footnote{After 
the first version of our paper had appeared on the web,
Chaudhuri \cite{AKC} posted a paper addressing similar issues.}

\begin{itemize}
\item[(i)] To establish a framework in which the influence of
fluctuations in $E_T$ on $\psi$ suppression can be calculated, to
derive a correction factor containing $E_T$ and to investigate the
limits under which it reduces to the {\it ad hoc} expression of eq.(\ref{eps}).
\item[(ii)] To investigate the importance of ``cross talk'', i.e. to
calculate the correlation function for the coincidence of hadrons in
different rapidity intervals. 
\end{itemize}

To deal with issue (i) 
we consider a nucleus-nucleus collision at fixed impact parameter
$b$. The individual $NN$ collisions produce many particles, possibly first
as strings or partons which then convert into the observed hadrons. 
In what follows
we will speak of particles, leaving open whether we deal with partons or 
hadrons. We consider the 6-dimensional phase-space for the particles
and divide it into cells
numbered as $i = 0,\, ...,\, k$, where the index $0$ is reserved for the
cell in which $\psi$ is found together with its comoving particles.
We denote by $m_i$ the number of particles in the $i$-th cell and by
$p_i(m_i)$ their probability distribution. It will be characterized by 
a mean value
${\bar m}_i$ and a variance $\sigma^2_i$. The probability to find $m_0$
particles in cell $0$ in an event with a total of $M$ produced 
particles is then
\be
\label{(8)}
\varphi(M,m_0) = \!\!\!\!\! \sum_{m_1,\ldots,m_k} \!\!\!
p_0(m_0)\ldots p_k(m_k)\,\delta(M-\sum^k_{i=0}m_k).
\ee
Note that we do not sum over $m_0$. We also
introduce the probability distribution
\be
\label{9a}
\phi_T(E_T,M) = \frac{e_T}{\sqrt{2\pi\tilde\sigma^2_{E_T}}}
\exp[\,- (E_T-e_TM)^2\,/\,2 \tilde\sigma^2_{E_T}],
\ee
which describes the correlation between the observed transverse energy
$E_T$ and the total number $M$ of particles, $e_T$ being the mean
transverse energy which each particle contributes to the total
$E_T$. 

The correlation between the observed transverse energy $E_T$ and the
number $m_0$ of particles, which suppress $\psi$ is then given by
\be
\label{(9)}
P(E_T,m_0) = \sum_{M}\, \phi_T(E_T,M)\ \varphi(M,m_0)\,.
\ee
If Gaussian distributions are used for the correlation functions $\phi_T$ and
$\varphi$ and if the sums over $M$ and $m_0$ are replaced by
integrals, eq.~(\ref{(9)}) can be evaluated exactly, but a rather 
cumbersome expression results. For the case of many cells, $k\gg 1$, 
the result, can be  factorized as
\be
\label{(10)}
P(E_T,m_0) = P_T(E_T,b)\ P_c(m_0,\bar m_0,E_T).
\ee
Here the distribution of the fluctuations in $E_T$ is
\be
\label{(11)}
P_T(E_T,b) = \frac{1}{\sqrt{2\pi\sigma^2_{E_T}}}\ \exp[\, - (E_T-\bar
E_T(b))^2/2
\sigma^2_{E_T}],
\ee
where $\bar E_T$ is the mean value of the produced transverse energy
for a given value of the impact parameter.
In eq.~(\ref{(11)}) $P_T$
does not depend on the number $m_0$ of particles in cell $0$ because they 
contribute a negligible amount. However,
the distribution function $P_c$ of particle number $m_0$ 
does depend on $E_T$ via
\bea
\label{(12)}
P_S (m_0,\bar m_0,E_T) &=& \frac{1}{\sqrt{2\pi\sigma_c^2}} \\
& & \!\!\!\!\!\!\! \times \, \exp[\,- (m_0-\bar m_0(\vec b,\vec s)
\hat\epsilon(E_T))^2/2\sigma^2_0]\,, \nonumber
\eea
where the dependence on transverse energy is contained in the expression
\be
\label{(13)}
\hat\epsilon(E_T)=1 + \alpha_0\ \frac{(E_T-\bar E_T(b))}{\bar E_T(b)},
\ee
which is close in shape to the one of eq.~(\ref{(1)}).
The factor 
\be
\label{(14)}
\alpha_0=\frac{\sigma^2_0}{\bar m_0}\cdot \frac{e_T\bar E_T}{\sigma^2_{E_T}}
\ee
depends on the mean value $\bar m_0$ and the variance
$\sigma_0$
 of the probability
distributions $p_0$ for the number of particles in cell $0$
and on the corresponding quantities $\bar E_0(b)$ and
$\sigma^2_{E_T}(b)$ for the distribution of 
 the observed transverse energy.
When both distributions are normal, i.e. $\bar m_0=\sigma_0^2,\
\sigma_{E_T}^2=e_T\bar E_T$,
 each of the two factors in
eq.~(\ref{(14)}) equals 1. Then $\alpha_0=1$ and
\be
\label{(17a)}
\hat\epsilon(E_T)=\epsilon(E_T)=\frac{E_T}{\bar E_T}
\ee
as assumed in refs. \cite{CFK,BDO}. A value for the second factor 
$e_T\bar E_T/\sigma^2_{E_T}$
in 
$\alpha_0$ can be deduced from a fit of
calculated $E_T$ distributions (using eq.~(\ref{(11)})) to the
measured one. One finds values between 1 and 0.7
\cite{CFK,BDO}.
 Although we have no direct information
on the ratio $\sigma^2_0/{\bar m_0}$, the multiplicity distribution of
produced particles in $pp$ collisions are known to be
negative binomials \cite{CARR} for
which $\sigma^2_0/\bar m_0=1+\frac{\bar m_0}{k}$, where $k$, the order
of the binomial, is found to be of order 3 or 4 for $pp$ collisions at
ISR energies. Therefore $\alpha_0$ is a product of two factors, one
larger than 1, the other smaller than 1. Although they may compensate each
other to a large extent, we are not in a position to give a reliable value
for $\alpha_0$. The derivation of the expression
$\hat\epsilon(E_T)$, eq.~(\ref{(13)})  is the first result of our
paper.

Now we address issue (ii), the effect of incomplete cross talk.
 In the
above derivation it has been tacitly assumed that produced particles 
in cell $0$,
which contribute to $\psi$ suppression, also contribute to the observed $E_T$.
As explained in the introduction,  the 
rapidity intervals for particles which suppress $\psi$ and for those which
produce  the observed $E_T$ do not overlap.
 Therefore the observables $m_0$ and $E_T$ should
not be correlated at all, unless there is cross talk between 
rapidity intervals.
String formation and decay is a possible mechanism which leads to correlations
between produced particles in different rapidity intervals.

The following calculation of cross talk is based on string formation
and their decay within the dual parton model \cite{CST} for particle
production (Fig.~1): Two protons, $P_1$ and $P_2$, with rapidities
$y=0$ and $y=Y$, respectively, interact via color exchange. After the
interaction, two strings form: string 1 between the quark with $y_1$
and the diquark of $P_2$ (assumed to have rapidity $Y$) and string 2
between the quark with $y_2$ from $P_2$ and the diquark from
$P_1$. The probability distribution for the occurrence of string $i$
is denoted by $w_i(y_i),\ i=1,2$ and is normalized to 1.
The strings fragment into hadrons where the number of produced hadrons
per rapidity interval is roughly independent of $y$. In a $NN$
collision, a hadron with rapidity $y_a$ can arise from each string,
provided it covers the rapidity $y_a$. Therefore the
probability to find a hadron with rapidity $y_a$ in the event shown in
Fig.~1 is
\be
\label{(18)}
P(y_a)= \!\int\! dy_1dy_2 \ w_1(y_1)\,w_2(y_2)\ 
[\theta(y_2-y_a)+\theta(y_a-y_1)]\,,
\ee
where each $\theta$-function refers to the contribution of one
string. Furthermore, the probability to observe two hadrons, one at
rapidity $y_a$ and another of $y_b$ is
\bea
\label{(19)}
P(y_a,y_b) &=& \int\! dy_1 dy_2\ w_1(y_1)\, w_2(y_2)\\
& &\!\!\!\!\!\!\!\!\!\!\!\! \times\, [\theta(y_2-y_a)\,\theta(y_2-y_b)
+ \theta (y_a-y_1)\,\theta(y_b-y_1)]\,. \nonumber
\eea
Let us identify $y_a$ with the rapidity of the charmonium $y_\psi$ and $y_b$
with the rapidity $y_{E_T}$ of a hadron contributing to the observed $E_T$. 
The probability $\alpha_x$ for cross talk  $(\alpha_x\leq 1)$ is then
given by the ratio
\be
\label{(20)}
\alpha_x(y_\psi,y_{E_T})=\frac{P(y_\psi,y_{E_T})}{P(y_\psi)}\,,
\ee
where $P(y_\psi,y_{E_T})$ gives the probability that a hadron at $y_\psi$
(comoving with the $\psi$) and a hadron  at $y_{E_T}$ (contributing to
$E_T$) are correlated and where $P(y_\psi)$ gives the probability to find
a comoving hadron at $y_\psi$ without any further condition.

The probability distribution $w_i(y_1)$ for the occurrence of the string
$i$ is taken proportional to the quark distribution function
$f_i(x_i)$ via $dx_i f_i(x_i) = dy_i w_i(y_i)$, where  $x_i$ is the
fractional momentum of the quark.
If we confine ourselves to the low values of $x_i$ then $f_i(x_i) \sim
x_i^{-\frac{1}{2}}$ and
\bea\label{(22)}
w_1(y_1)&=& c\, e^{-\frac{1}{2} y_1},\nonumber\\
w_2(y_2)&=& c\, e^{-\frac{1}{2}(Y-y_2)},\eea
where the normalization drops in the ratio   eq.~(\ref{(19)}). In the
experiment under consideration, we are interested in the cross talk
between the comoving hadrons $(3\leq y_\psi \leq 4)$ and the hadrons in
transverse energy $E_T (1\leq y_{E_T} \leq 2.3)$. Instead of integrating
$\alpha_x$ over these intervals, we evaluate  eq.~(\ref{(20)}) at the
mean values $y_\psi=3.5,\ y_{E_T}=1.65$. A straightforward calculation then
leads to
\be
\label{(23)}
\alpha_x(3.5,1.65)=0.82\,.
\ee
The high value of the coefficient for cross talk ($\alpha_x=1$
corresponds to complete cross talk) reflects the property of  the probability
distributions $w_i$ in that they are largest for those strings
which span the full rapidity range. In our calculation, we have
neglected the strings involving sea quarks. Since these strings are
shorter, their inclusion would reduce the value of $\alpha_x$.

With this result, the factor $\hat\epsilon(E_T)$ from
eq.~(\ref{(13)}) has to be modified to
\be\label{(24)}
\hat\epsilon(E_T)=1+\alpha_0\cdot\alpha_x\left(\frac{E_T-\bar
E_T(b)}{\bar E_T(b)}\right)\ee
with $\alpha_0$ from  eq.~(\ref{(14)}) and $\alpha_x$ from 
 eq.~(\ref{(20)}).  Eq.~(\ref{(24)}) is the final result of our paper.

We now look at the experimental $J/\psi$ suppression as a function of 
$E_T$ for $Pb\!+\!Pb$ collisions at 158$\,A$ GeV and compare with it the 
results of several calculations in order to show the importance of 
the effects discussed in this paper (Fig. 2).
 We have repeated the calculation of
\cite{BDO} with the form given in eq.~(\ref{(7)})  for the anomalous
suppression.  However, the expression $\epsilon=E_T/\bar E_T$ in front of
$n_p(\vec b,\vec s)$ is replaced by the expression $\hat\epsilon$
 derived in 
eq.~(\ref{(24)}) with $\alpha=\alpha_0\cdot\alpha_x$ which describes
the influence
of the shape of the probability distributions and of
 the incomplete cross talk. Then $\alpha=0$ means that 
no fluctuations are taken into account while $\alpha =1$ gives the maximal
influence of fluctuations as in \cite{CFK,BDO}. 
A reasonable estimate may be $\alpha=0.8$, but we also give a curve
for $\alpha=0.4$.
 While $\alpha=1$ and $\alpha=0.8$ are compatible with the data, 
$\alpha=0.4$ and smaller values are definitely ruled out, provided the
effect discussed in ref. \cite{CKS} is not too important. 

In this paper we have derived and discussed the phenomenological 
prescriptions by Capella
{\it et al.} \cite{CFK} and Blaizot {\it et al.} \cite{BDO} to
calculate  $J/\psi$ suppression in the region of large values of $E_T$
where the transverse energy fluctuates. The derivation has shown
which physical parameters determine the strength with which
fluctuations in $(E_T-\bar E_T)/\bar E_T$ influence $J/\psi$
suppression. These are the width parameters for the multiplicity
distributions of comovers and  for the $E_T$ distributions and
the correlation function for  incomplete cross talk. We conclude that
the {\it ad hoc} factor, eq.~(\ref{(1)}), holds to a good approximation.

\vspace{0.3in}

\noindent{\bf Acknowledegement:} This work has been supported in part by the
German federal ministry BMBF under contract number 06 HD 642.

\newpage

\begin{figure}[ht]
\centerline{\psfig{figure=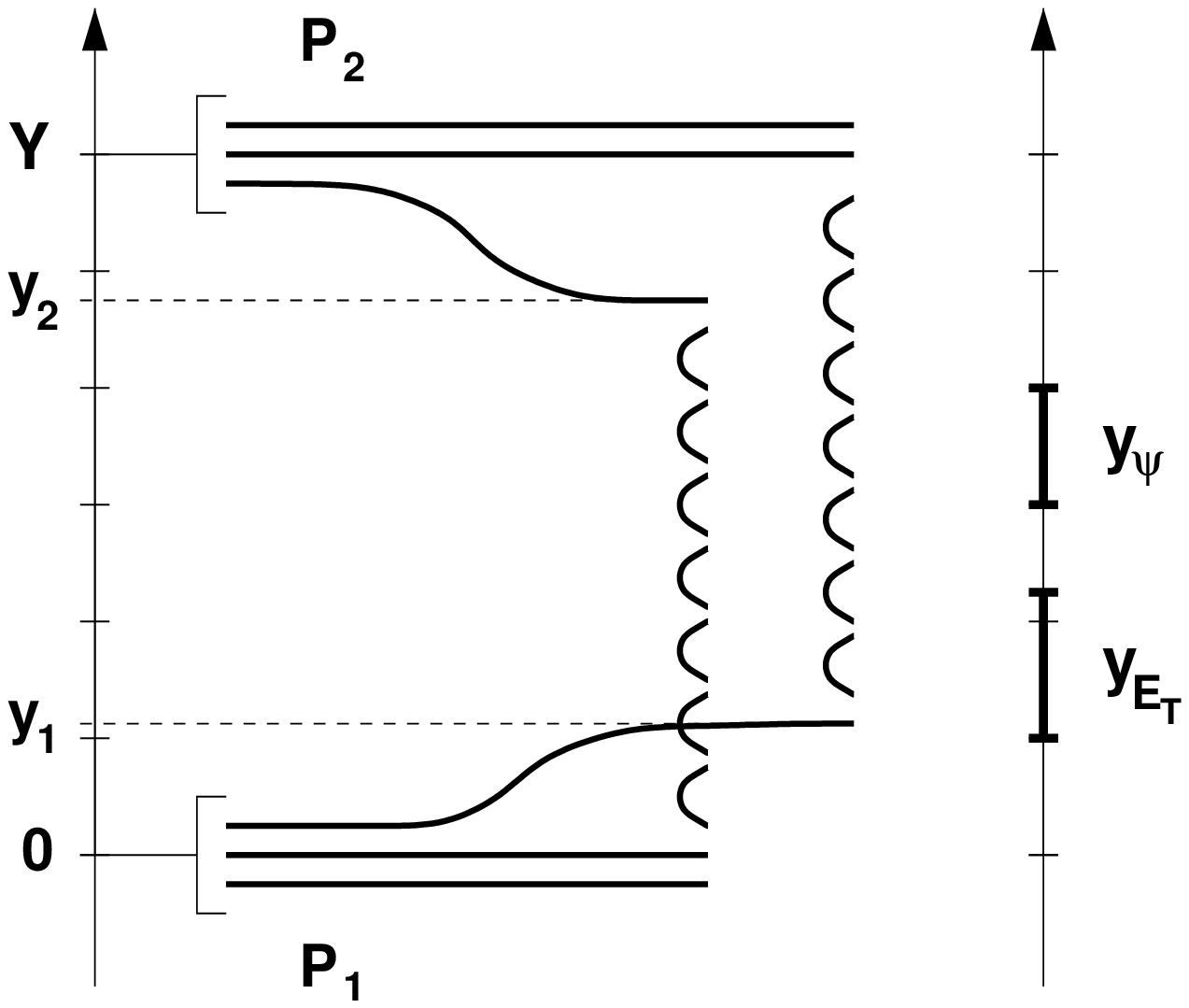,width=12cm,angle=-90}}
\protect\caption{
\setlength{\baselineskip}{16pt}
Particle production for proton-proton collisions
within the dual parton model \cite{CST}. Two protons, $P_1$ and $P_2$, with
rapidities $y=0$ and $y=Y$ collide and exchange color
after which two strings form. We study production of hadrons at
the rapidities $y_\psi$ (which comove with the $\psi$)
and $y_{E_T}$ (which contribute to the transverse energy $E_T$).
}
\label{figure1}
\end{figure}
\begin{figure}[hb]
\centerline{\psfig{figure=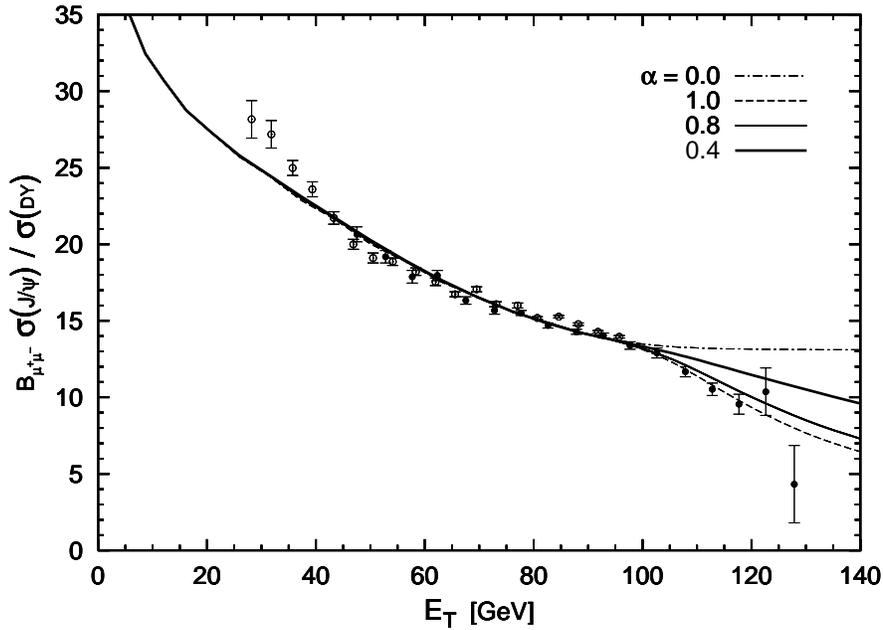,width=12cm}}
\protect\caption{
\setlength{\baselineskip}{16pt}
The experimental values for the ratio $J/\psi/
DY$ in $Pb+Pb$ collisions at 158 $A$ GeV \cite{NA50}. The curves are
calculated like in the paper by Blaizot {\it et al.} \cite{BDO} and
differ by the degree $\alpha$ to which fluctuations are taken into
account $\alpha=\alpha_0\cdot\alpha_x$, eq.~(\ref{(23)}). $\alpha=0:$
Fluctuations are not accounted for. $\alpha=1:$ Fluctuations influence
suppression fully. $\alpha=0.8$ and $\alpha=0.4:$ Partial reduction of
the importance of fluctuations.
}
\label{figure2}
\end{figure}
\end{document}